\newlength\smallfigwidth
\documentclass[twocolumn,spanish,aps,prb,showpacs,superscriptaddress,floats,amsmath,amssymb,floatfix]{revtex4}

\usepackage{color}
\usepackage{amsmath}
\usepackage{graphicx}
\usepackage{graphics}
\usepackage{pdftexcmds}
\usepackage{ifpdf}
\usepackage{amssymb}
\usepackage[utf8]{inputenc}
\usepackage{tikz}
\usepackage{multirow}
\usepackage{comment}
\makeatletter
\@ifundefined{textcolor}{}

{
 \definecolor{BLACK}{gray}{0}
 \definecolor{WHITE}{gray}{1}
 \definecolor{RED}{rgb}{1,0,0}
 \definecolor{GREEN}{rgb}{0,1,0}
 \definecolor{BLUE}{rgb}{0,0,1}
 \definecolor{CYAN}{cmyk}{1,0,0,0}
 \definecolor{MAGENTA}{cmyk}{0,1,0,0}
 \definecolor{YELLOW}{cmyk}{0,0,1,0}
 }
 
\definecolor{J1}{rgb}{0.0, 0.5, 1.0}
\definecolor{Jx}{rgb}{0.6, 0.33, 0.73}
\definecolor{Jp}{rgb}{0.28, 0.24, 0.2}
\definecolor{bond}{rgb}{0.2, 0.2, 0.9}
\allowdisplaybreaks
\makeatother
\def \ba {\begin{eqnarray}}
\def \ea {\end{eqnarray}}

\begin{document}
\title{Field-induced pseudo-skyrmion phase in the antiferromagnetic kagome lattice}
\author{M. E. Villalba}
 \email{mvillalba@iflysib.unlp.edu.ar}
\affiliation{Instituto de F\'isica de L\'iquidos y Sistemas Biol\'ogicos, CONICET, Facultad de Ciencias Exactas, Universidad Nacional de La Plata, 
1900 La Plata, Argentina;
   Departamento de F\'{i}sica, FCE, UNLP, La Plata, Argentina}
\author{F. A. G\'omez Albarrac\'in}
\affiliation{Instituto de F\'isica de L\'iquidos y Sistemas Biol\'ogicos, CONICET, Facultad de Ciencias Exactas, Universidad Nacional de La Plata, 
1900 La Plata, Argentina;
   Departamento de F\'{i}sica, FCE, UNLP, La Plata, Argentina}
   \affiliation{Departamento de Ciencias B\'asicas, Facultad de Ingenier\'ia, UNLP, La Plata, Argentina}
   \author{H. D. Rosales}
\affiliation{Instituto de F\'isica de L\'iquidos y Sistemas Biol\'ogicos, CONICET, Facultad de Ciencias Exactas, Universidad Nacional de La Plata, 
1900 La Plata, Argentina;
   Departamento de F\'{i}sica, FCE, UNLP, La Plata, Argentina}
    \affiliation{Departamento de Ciencias B\'asicas, Facultad de Ingenier\'ia, UNLP, La Plata, Argentina}
   \author{D. C. Cabra}
\affiliation{Instituto de F\'isica de L\'iquidos y Sistemas Biol\'ogicos, CONICET, Facultad de Ciencias Exactas, Universidad Nacional de La Plata, 
1900 La Plata, Argentina;
   Departamento de F\'{i}sica, FCE, UNLP, La Plata, Argentina}
   \affiliation{Abdus Salam International Centre for Theoretical Physics, Associate Scheme, Strada Costiera 11, 34151, Trieste, Italy}
   \begin{abstract}

We study the effects of an in-plane Dzyaloshinskii-Moriya interaction under an external magnetic field in the highly frustrated kagome antiferromagnet. We focus on the low-temperature phase diagram, which we obtain through extensive Monte-Carlo simulations. We show that,
given the geometric frustration of the lattice, highly non trivial phases emerge. 
At low fields, lowering the temperature from a cooperative paramagnet phase, the kagome elementary plaquettes form non-coplanar arrangements with non-zero chirality, retaining a partial degeneracy.
As the field increases, there is a transition from this ``locally chiral phase'' to an interpenetrated spiral phase  with broken $\mathcal{Z}_{3}$ symmetry. 
Furthermore, we identify a quasi-skyrmion phase in a large portion of the magnetic phase diagram, which  we characterize with a topological order parameter, the scalar chirality by triangular sublattice.  This pseudo-skyrmion phase (pSkX) consists of a crystal arrangement of three interpenetrated non-Bravais lattices of skyrmion-like textures, but with a non-(fully)-polarized core.  The edges of these pseudo-skyrmions remain polarized with the field, as the cores are progressively canted.
Results show that this pseudo-skyrmion phase is stable up to the lowest simulated temperatures, and for a broad range of magnetic fields.
\end{abstract}


\maketitle 

\section{Introduction}

Magnetic skyrmions are topological vortex-like spin structures where the spins point in all directions wrapping a sphere \cite{SkyrmionFirst}. 
In particular, in the last years, the Skyrmion crystal (SkX) phases have triggered a huge interest because of their important role in the 
electronic transport in conection with technological application devices \cite{TechDev}. The most simple situation where such SkXs are 
stabilized correponds to the ferromagnetic systems in a magnetic field including  Dzyaloshinskii-Moriya (DM) interactions \cite{bogda1,bogda2,bogda3,bogda4,bogda5,bogda6,bogda7,bogda8}. 
Also, it has been shown in numerous works that the SkX's can be induced by competing interactions in ferromagnetic and mixed ferro/antiferro-magnetic systems 
\cite{17,DMSkx}. Finally, the presence of local anisotropies can stabilize different skyrmion-like crystal phases under a magnetic field, which
lead to merons-like structures in metastable states\cite{12}.

In this direction, the search for new systems with skyrmions phases in a wide range of magnetic field and temperature is an central issue in the field of topological magnetic materials. 
One ingredient that may play a central role in this topic is the magnetic frustration, which in many cases, is associated with exotic spin orders having non-collinear or non-coplanar spin structures. 

Recently, the emergence of skyrmion textures has been actively explored in frustrated lattices \cite{naga,1,osorio1,osorio2,Yu2018,Loss2019}. In fact, in a previous work (see Ref. [\onlinecite{1}]), some of the authors (see also Ref. [\onlinecite{Loss2019}]) have shown that in the antiferromagnetic triangular lattice the competition between nearest neighbor exchange couplings and an in-plane DM interaction gives rise to a low temperature stable topological phase for a range of magnetic fields. This phase is characterized by three interpenetrated skyrmion crystals, one by sublattice. 

In this context, the highly frustrated kagome antiferromagnet  provides an alternative arena for studying emergent phenomena in magnets of strong frustration.  A crucial point of the  antiferromagnetic kagome lattice is its high degeneracy. This feature, combined with  the chiral anisotropy induced by the DM interaction could induce different types of  topologically non trivial phases. In the last years, 
materials with in- and out-of-plane DM interactions with an antiferromagnetic  kagome structure have been thoroughly studied, for example by P. Mendels et al (Herbermisthite)\cite{kagomeDM1}, B. Canals et al (Fe- and Cr-based jarosites, etc.)\cite{kagomeDM2} (for a review see Ref.[\onlinecite{NormanReview}]). 
Last but not least, the possibility to generate a DM interaction in ultrathin films with perpendicular magnetic anisotropy in multilayer structures leads to the emergence of interfacial non-collinear spin textures (skyrmions and chiral domain walls) induced by DM interactions in such magnetic films \cite{Fertetal,kagomeDM4}.

A key question that arises is what is the role of the magnetic frustration and hugh degeneracy in the formation of skyrmion spin textures. Motivated by this, we consider the inclusion of a specific DM interaction in the pure antiferromagnetic Heisenberg model on the kagome lattice and study the consequences of the combination of high degeneracy, thermal fluctuations and anisotropic interactions. In particular, we explore the possibility of skyrmion-like textures in the proposed model.

We show that, under the action of an  external magnetic field,  there are a number of different exotic low-temperature phases: at low non-zero magnetic field  a phase without global order, reminiscent of the pure Heisenberg model in the kagome lattice,  but formed by clusters with non-zero local chirality, is stabilized. Then increasing the magnetic field, it leads to a tree-sublattice order with broken sublattice symmetry. For a larger magnetic field, a three sublattice pseudo-skyrmion crystal (pSkX) structure  is established with the particularity that the hidden pSkX magnetic order appear in a non-Bravais sublattice. The emergent pseudo-skyrmion unit structures do not fully wrap the sphere, but can be distinguished with the help of another topological parameter, the sublattice chirality. 
These pseudo-skyrmion structures have a remarkable feature: instead of being the result of overlapping skyrmions, with a fully polarized core and a radius 
smaller than the separation between them, as in \cite{12}, here the rims are completely polarized in the direction of the external field while the cores are not. 
In fact, these cores get canted as the field increases. 

The rest of the manuscript is organized as follows: in Section \ref{sec:model} we present and discuss the Heisenberg model on the kagome lattice including the DM interaction and the Zeeman coupling to an external magnetic field.
In Sec. \ref{sec:MC} we study the proposed model through extensive Monte Carlo simulations and examine the different phases stabilized by the Zeeman coupling, focusing on the topological pSkX.
Conclusions are presented in Sec. \ref{sec:conclusions}.

\section{Model}
\label{sec:model}

\begin{figure}[htb]
\centering
\includegraphics[width=0.7\columnwidth]{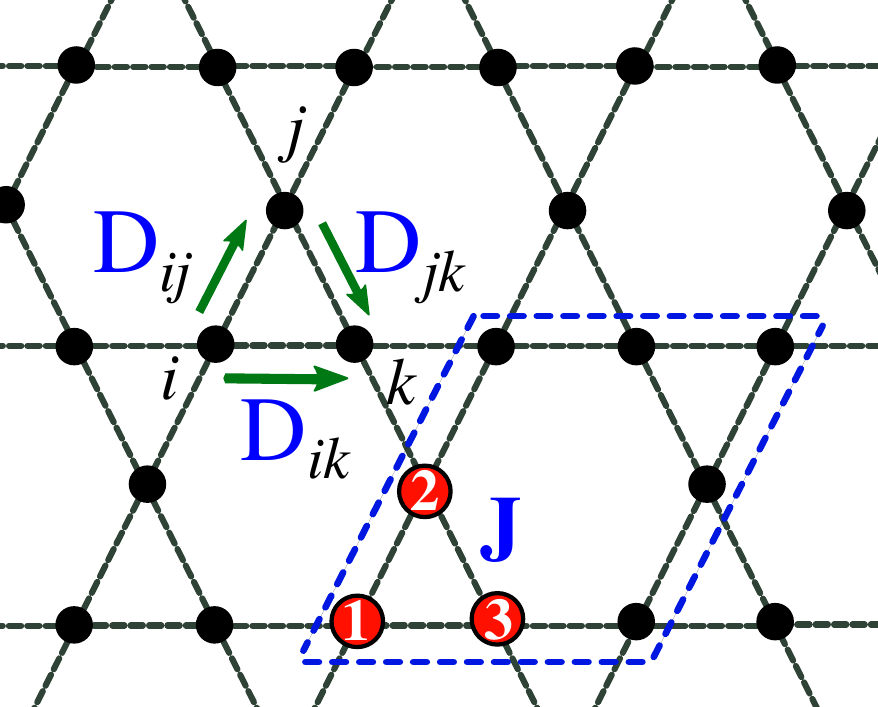}
\caption{\label{fig:latt} (Color online) kagome lattice.  The labels $1,2,3$ indicate the three  sublattices (dashed blue line indicates the unit cell). Small (green) arrows are Dzyaloshinskii-Moriya vectors $\bold{D}_{ij}$, $\bold{D}_{jk}$ and $\bold{D}_{ik}$ involved in the sites (labeled as) $i,j,k$.}
\end{figure}

We consider an antiferromagnetic Heisenberg model on the kagome lattice, with in-plane DM interaction,  immersed in a magnetic field.  The Hamiltonian is given by 
\begin{equation}
\label{eq:H}
H = J\sum_{\langle i,j\rangle}\mathbf{S}_{i} \cdot \mathbf{S}_{j} + \mathbf{D}_{ij}\cdot(\mathbf{S}_{i} \times \mathbf{S}_{j})- h \sum_{j} S_{j}^{z}
\end{equation}

\noindent where the magnetic moments $\mathbf{S}_{i}$ are three-component classical unit vectors at site ${\mathbf r}_i$, $\langle i,j \rangle$  indicates the sum over nearest neighbor sites, and $J>0$ is the antiferromagnetic exchange coupling. The DM interaction is defined by $\bold{D}_{ij}= D\, \delta \hat{r}_{ij}$, 
where $\delta\hat{r}_{ij}=(\mathbf{r}_{i}-\mathbf{r}_{j})/|\mathbf{r}_{i}-\mathbf{r}_{j}|$ is a unitary vector pointing along the nearest-neighbor bonds as shown in Fig. \ref{fig:latt}. This implies that this interaction is constrained
to the kagome plane, perpendicular to the external magnetic field $\bold{h}=h\hat{z}$. 
This choice of DM interaction proves to be adequate to develop Skyrmion phases in both ferromagnetic and antiferromagnetic systems \cite{1}. In this work, without loss of generality, we fix $D/J=0.2$, a value of $D/J$ that induces magnetic structures with sizes compatible with the systems size of the simulations.

The case $D = 0$ has been widely studied in the last decades \cite{2,3,4,5,6,7,19,21,9}. It is well known that the antiferromagnetic Heisenberg model for classical spins in the kagome lattice presents a rich phenomenology due to its high degeneracy. 
In zero field, magnetic moments form a infinitely degenerate $120^{\circ}$ spin-structure. 
Over the last few years, much effort has been devoted to study the mechanisms or interactions that can lift this degeneracy and the consequent emergence of non-trivial phases. One of these well known effects is that due to the inclusion of thermal fluctuations, the system goes from a paramagnetic (at high temperaure) to a cooperative paramagnetic  phase  
\cite{3,7}; while at low temperature the order-by-disorder mechanism selects a submanifold of coplanar states. A magnetic field partially relieves this degeneracy, and state selection by thermal fluctuations is still at play. Thermal fluctuations stabilize two coplanar states at finite fields with different symmetries. At very low temperature, each type of coplanar state can be studied through multipolar order parameters \cite{7}. The inclusion of further neighbor exchange couplings can select and induce different magnetic orders   \cite{8,fundamentales,14}. Furthermore, the addition of an out-of-plane DM interaction favors a $q=0$ non-coplanar state \cite{18}

Due to the competition between the antiferromagnetic exchange $J$ which favors the coplanar configurations, and the in-plane DM interaction which favors the helical phases, we expect that the combination of these to terms results in a rich variety of chiral configurations which will be presented in the next section. 

\section{Monte Carlo Simulations And Phase Diagram}
\label{sec:MC}

To explore the low temperature behavior of the model presented in the previous section, we resort to Monte Carlo simulations. We use a combination of the
Metropolis algorithm and the overrelaxation method, doing microcanonical updates, and lowering the temperature in an annealing scheme. 
We performed our simulations in $3\times L^2$ site clusters, $L=36-60$, with periodic boundary conditions.
$10^5-10^6$ Monte Carlo steps (MCS) were used for initial 
relaxation, and measurements were taken in twice as much MCS. 

As a first approach to identify and characterize the different low temperature phases, 
we first inspect the standard quantities: namely, specific heat $C_{v}=\frac{\langle E^{2} \rangle - \langle E \rangle^{2}}{NT^{2}}$, magnetization  $M=\frac{1}{N} \left \langle \sum_{i} S_{i}^{z} \right \rangle $, absolute value of the magnetization $|M|=\frac{1}{N} \left \langle \sum_{i} |S_{i}^{z}| \right \rangle $ and magnetic susceptibility $\chi_{M}=\left \langle \frac{dM}{dh}  \right \rangle$.

\begin{figure}
\includegraphics[width=1.0\columnwidth]{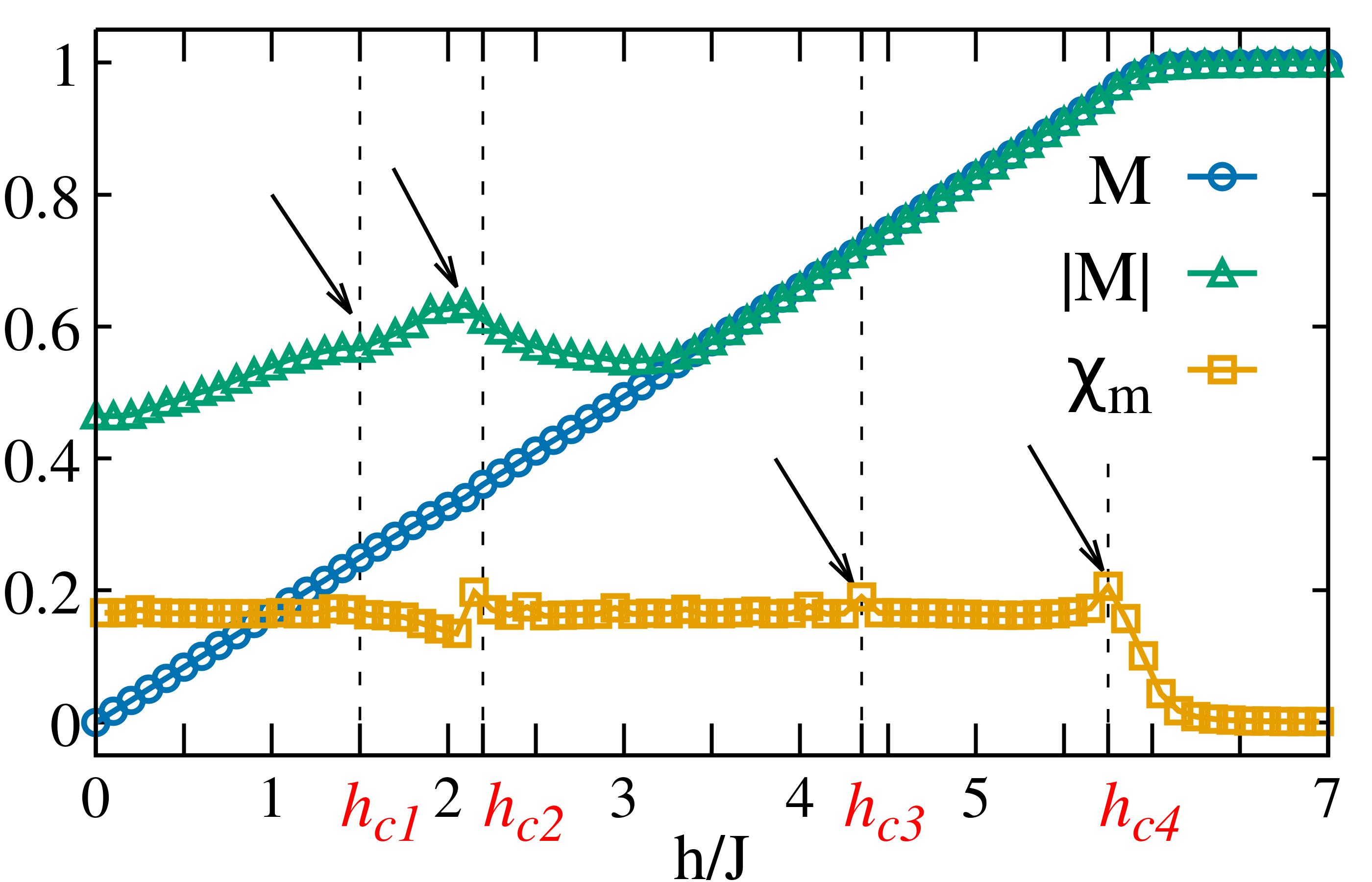}
\caption{\label{fig:magH} (Color online) Magnetization curves vs external field $h/J$ for $L=60$ lattice size,  at $T/J=2\times10^{-2}$. Average Magnetization $\langle M \rangle$ (blue open circles),  Magnetization modulus $|M|$ (green open triangles), susceptibility $\chi_M= dM/dh $ (yellow open squares). The black arrows indicate four features in these curves. The value of the fields where these feature emerge, the critical fields, $h_{c1},h_{c2},h_{c3}$, are indicated by dashed lines. The critical field $h_{c4}$ correspond to the saturation field where all the spins are polarized.} 
\end{figure}

In Fig. \ref{fig:magH} we show typical curves of magnetization $M$, its absolute value $|M|$ and the susceptibility $\chi_{M}$ as a function of the magnetic field at $T/J=2\times 10^{-3}$.
We can identify four features in these curves, indicated by vertical arrows
in the figure: a bump in $|M|$ at $h_{c1}/J\sim 1.5 $, a peak in the susceptibility which matches a change in the behavior of $|M|$ at $h_{c2}/J\sim 2.1$ and a second peak in $\chi_{M}$ at $h_{c3}/J\sim 4.4$. The last feature at the critical field $h_{c4}\sim 5.7$, indicates the transition to the state completely polarized with the magnetic field. To obtain valuable information on the nature of each phase we compute the static spin structure factor $S_{\bf q}$ in the reciprocal lattice to identify the Bragg peaks that characterize the different spin-textures.

\begin{figure*}[ht]
\includegraphics[width=1\textwidth]{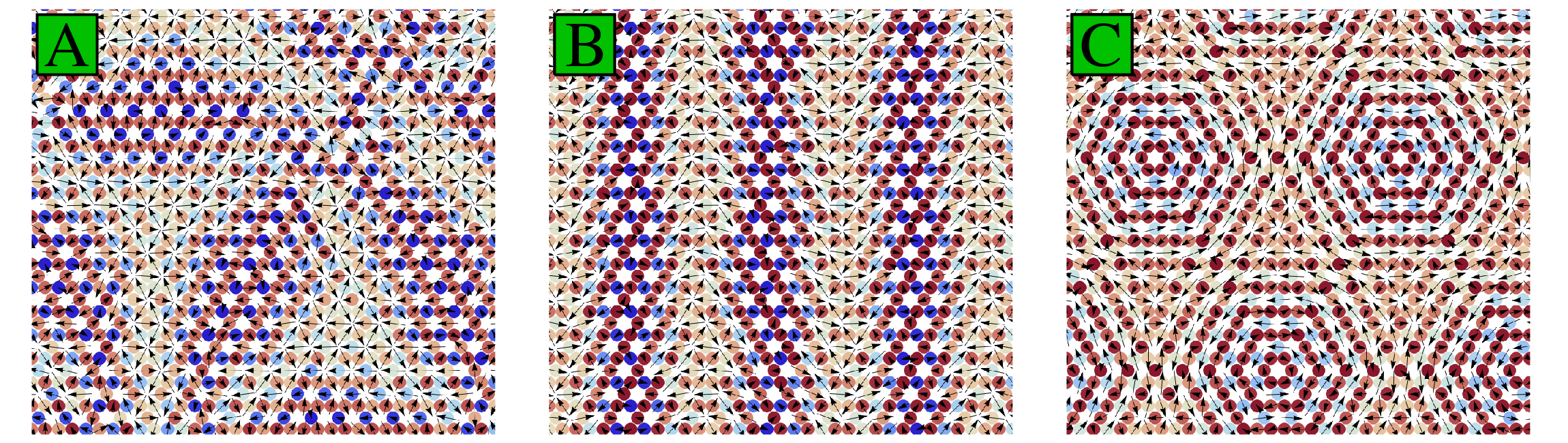}
\includegraphics[width=1\textwidth]{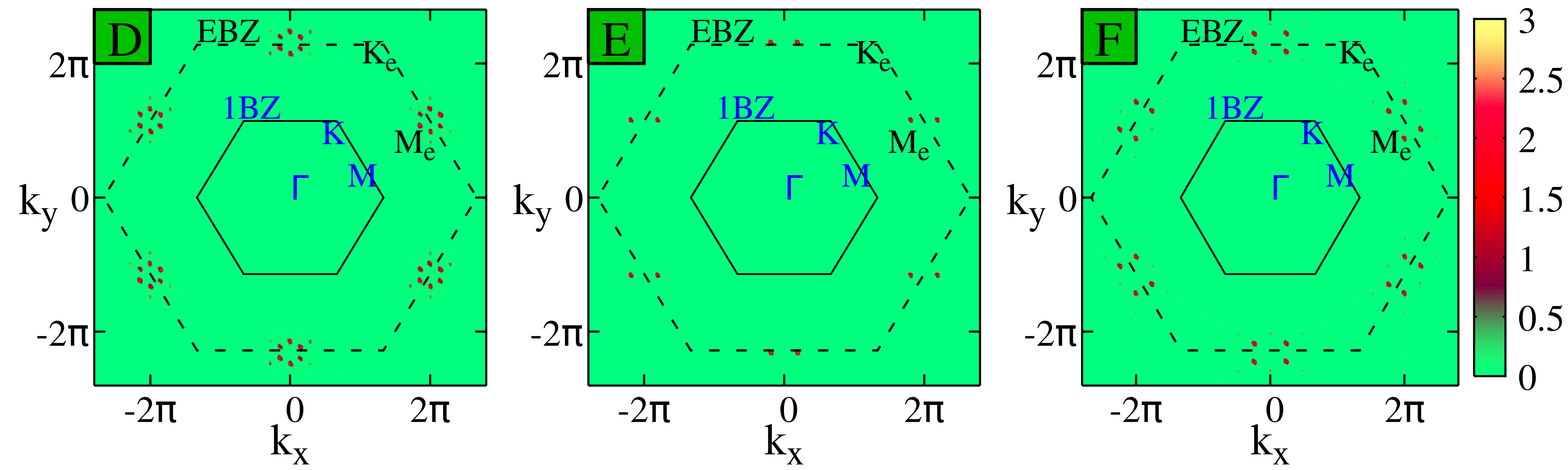}
\caption{ (Color online) Snapshots of the spin textures and the corresponding structure factors for $L=42$, $D/J=0.2$, $T/J=2\times 10^{-3}$. As the external magnetic field increases different 
structures can be identified. For this plot we select one $h/J$ value in representation of each phase:  the locally chiral phase at $h/J=0.8$ (A and D panels); interpenetrated spirals at $h/J=1.8$ (B and E panels), and pseudo-skyrmion crystal at  $h/J=2.8$ (C and F panels).}
\label{fig:snaps}	
\end{figure*}

The components $S_{\bf q}^{\perp} $ and $S_{\bf q}^{\parallel}$, perpendicular and parallel to the external field respectively, are defined as:

\begin{eqnarray}
S_{\bf q}^{\perp}&=&\frac{1}{N} \langle |\sum_{i} S_{i}^{x} \ e^{i {\bf q}\cdot {\bf r}_i} |^2 + |\sum_{i} S_{i}^{y} \ e^{i {\bf q} \cdot {\bf r}_i} |^2\rangle \label{factesperp}\\
S_{\bf q}^{\parallel}&=& \frac{1}{N} \langle |\sum_{i} S_{i}^{z} \ e^{i{\bf q}\cdot {\bf r}_i} |^2\rangle\label{factespara}
\end{eqnarray}

In Fig. \ref{fig:snaps} we show representative snapshots (top) and the corresponding structure factors $S_{\bf q}^{\perp}$ (bottom) of the low temperature phases as a function of the external magnetic field. In the $S_{\bf q}^{\perp}$ plots,  the first Brillouin zone (1BZ drawn with solid lines) and  the extended Brillouin zone (EBZ drawn with dashed lines) are indicated. By inspection of Fig. \ref{fig:snaps} we find:
\begin{itemize}
\item{For very low magnetic fields $h<h_{c1}$, the magnetic structure retains some of the degeneracy present for the case $D = 0$ and $h = 0$. From a typical snapshot, it can be seen that
elementary triangles form out of plane structures. Six bright peaks emerge around every high-symmetry point $\bf{M_e}$ in the spin structure factor as is shown in Fig. \ref{fig:snaps}(D). Half of these points  (18 in total)  are  inside the extended Brillouin zone (EBZ).
}
\item{For slightly higher fields $h_{c1}<h<h_{c2}$, the Zeeman coupling induces a striped/spiral-like structure, with single-q peaks in the $\bf{M_e}$ region of the EBZ, Fig. \ref{fig:snaps}(E).}

\item{In a broad region of intermediate magnetic fields $h_{c2}<h<h_{c3}$ a non trivial swirling structure emerges (see Fig. \ref{fig:snaps}(C)). Visually, it is reminiscent of the interpenetrated skyrmion phase AF-SkX found in the  triangular antiferromagnetic lattice\cite{1}.
In the structure factor, 12 peaks emerge in the EBZ, indicating a triple-q structure, which may be a hint of a hidden skyrmion-like texture.}
\end{itemize}
Now, we proceed to further explore and characterize in detail these low temperature phases.

\subsection{Lower $h/J$ multi-q states}

%
\begin{figure}[ht]
\centering
\includegraphics[width=1\columnwidth]{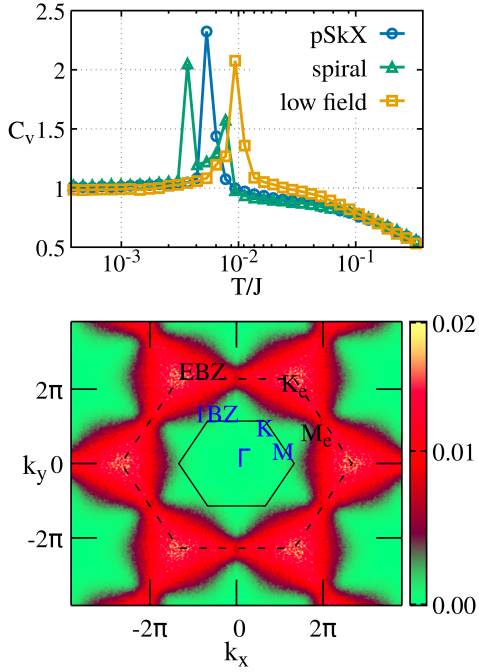}
\caption{(Color online) Top: Specific heat as a function of temperature for: low magnetic field at $h/J=0.5$ (yellow open squares), spiral at $h/J=1.6$ (green open triangles) and pSkX at $h/J=2.7 $ (blue open circles). Bottom: structure factor obtained from Monte Carlo simulations at temperature $T/J=8\times 10^{-2}$, $h/J=0.5$, for  $D/J=0.2$.}
\label{fig:SqLowH}
\end{figure}

At low external field, $h<h_{c1}$, there is an interesting behavior of the system with temperature. At  $T/J>0.03$, the system seems to be in a copperative paramagnet (CP) phase. This is illustrated in the structure factor, presented in the bottom panel of figure Fig. \ref{fig:SqLowH} showing similar  behaviour to that obtained for the pure kagome antiferromagnet in the CP
phase \cite{2}. It is characterized by the presence of ``pinch points'' in the $\mathbf{M_e}$ points of the EBZ, which are the signature of a classical algebraic spin liquid \cite{10,11}. The  CP and low temperature phases are separated by a phase transition at $T/J\approx 0.03$, where the specific heat exhibits a peak (top panel of Fig. \ref{fig:SqLowH}). 

\begin{figure}[hbt!]
\includegraphics[width=0.98\columnwidth]{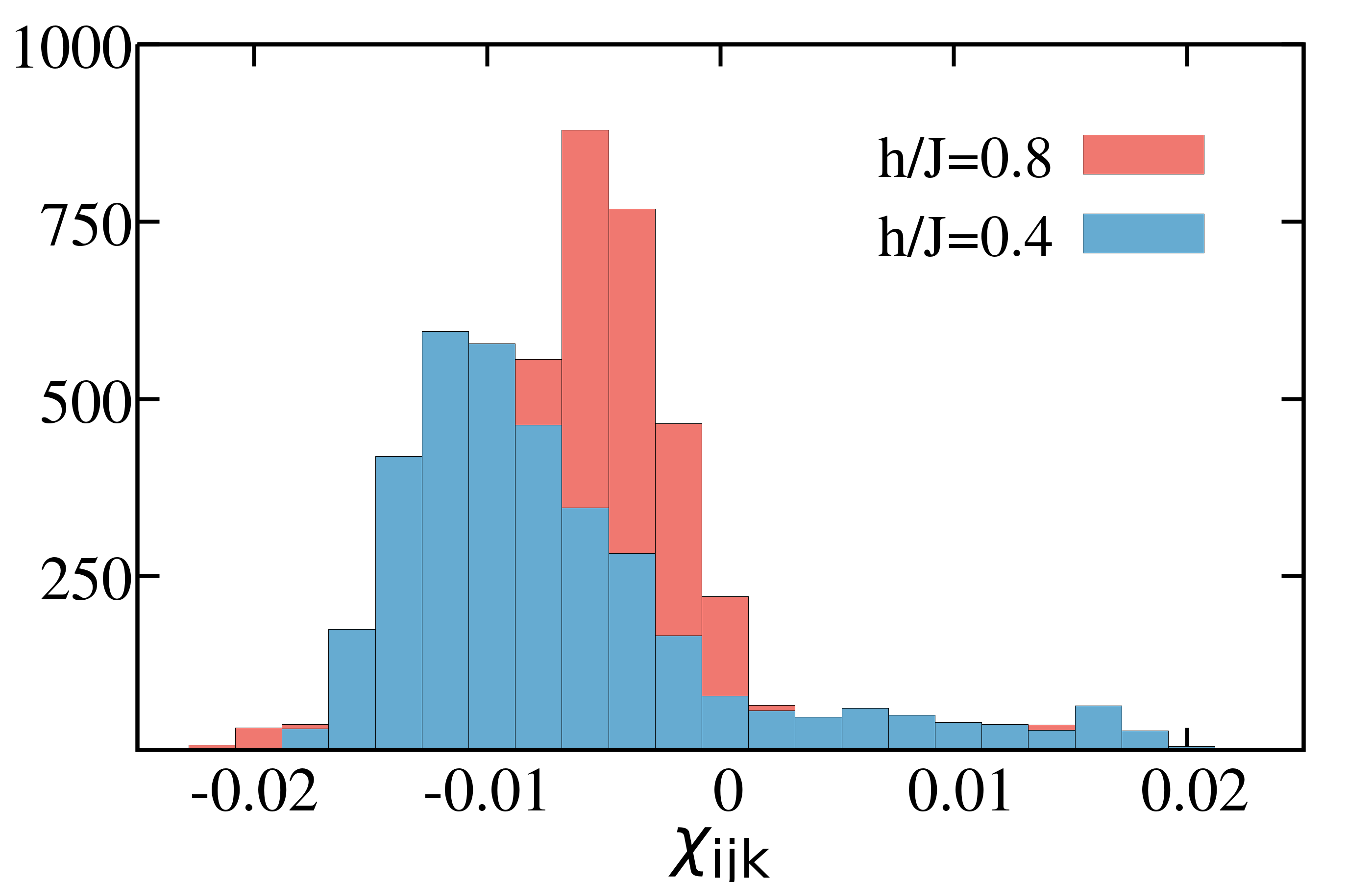}
\caption{ (Color online) Histogram of the nearest neighbor chirality $\chi_{ijk}$ per triangular plaquette, for two snapshots at $T/J=2 \times 10^{-3}$, $L=60$, $h/J=0.8$ (red) and $h/J=0.4$ (blue). }
\label{fig:histo}
\end{figure}

However, these low temperature phases at low magnetic field do not show a clear periodic magnetic structure. As we mentioned before, although no clear order is seen, it is evident that there are numerous unit triangles where the spins are arranged in a non-coplanar way, with different
orientations. This is most clearly shown inspecting the nearest neighbor scalar chirality per plaquette $\chi_{ijk}$  defined as:

\begin{eqnarray}
\chi_{ijk}&=&\bold{S}_{i}\cdot( \bold{S}_{j} \times \bold{S}_{k}) 
\label{eq:local_chira}
\end{eqnarray}

\noindent where labels $i,j,k$ indicate the positions ${\bf r}_i$, ${\bf r}_j$, ${\bf r}_k$ of each of the three spins of every elementary triangular plaquette of the triangular lattice.
In order to analyze the local distribution of the nearest neighbor scalar chirality in the plaquettes, we plot a histogram of the local values of $\chi_{ijk}$ obtained from snapshots of a $L=60$ lattice, at $T/J=2 \times 10^{-3}$, for $h/J=0.4,0.8$, in  Fig. \ref{fig:histo}. For a perfect translational invariant chiral state, a strong peak at a given value of $\chi_{ijk}$ is expected.  However, in this phase the values of $\chi_{ijk}$ are widely spread, confirming  the non-coplanar nature of the low temperature phases at low fields.

\begin{figure}[h!]
\includegraphics[width=0.5\textwidth]{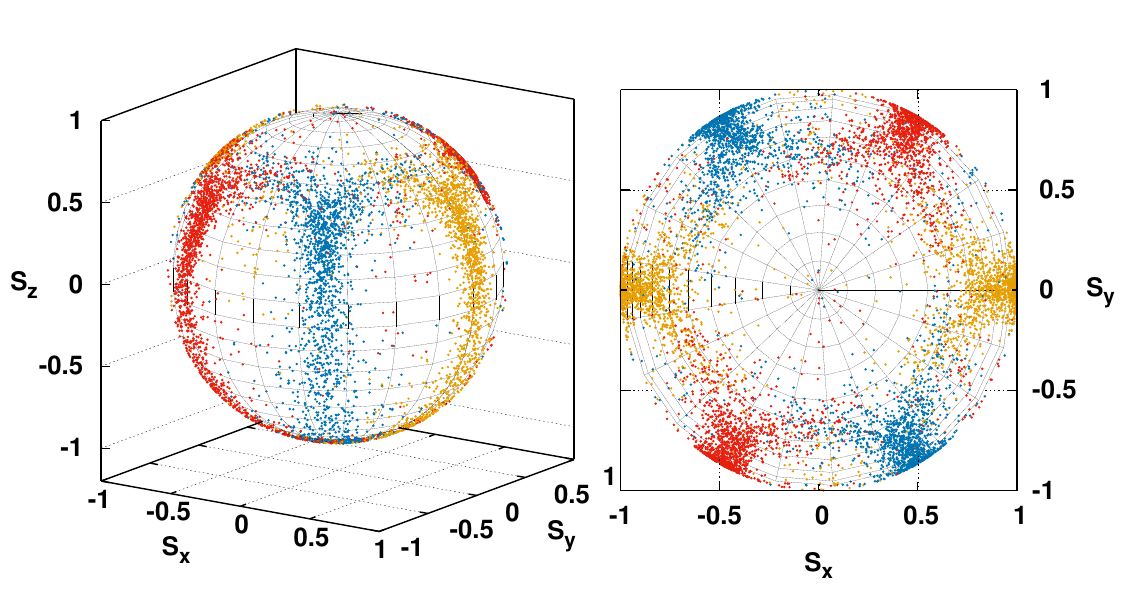}
\includegraphics[width=0.5\textwidth]{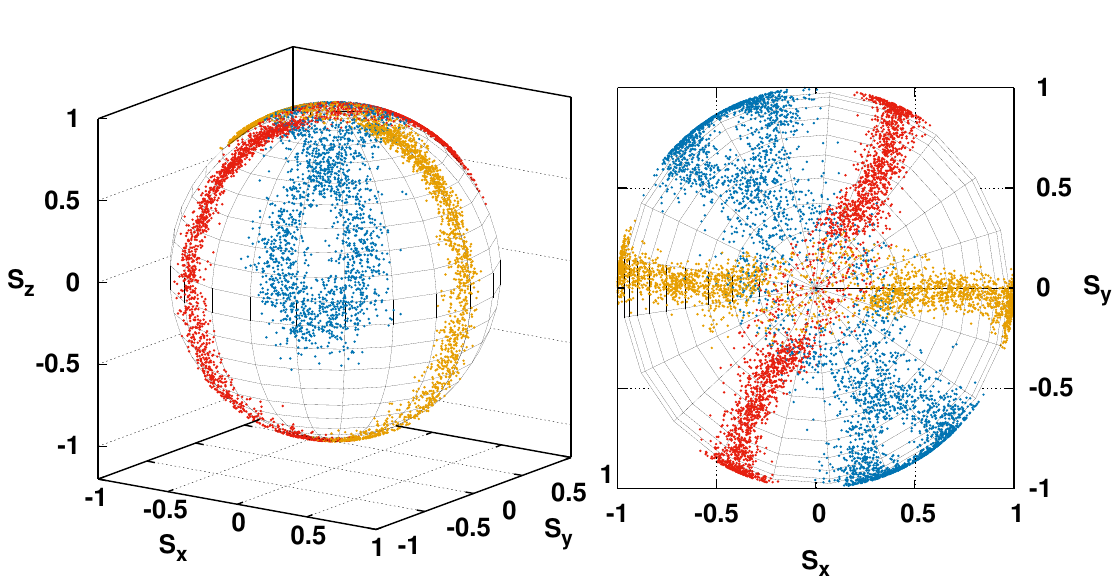}
\caption{\label{fig:sphesnaplowh} (Color online) Spherical snapshot at $T/J=2\times 10^{-3}$, for ``locally chiral'' phase at $h/J=0.8$ (top panel) and the spiral phase at $h/J=1.8$ (bottom panel).  Each color indicates a different triangular sublattice. The right column is the top view of the spherical snapshot. 
}
\end{figure}

Due to the lack of periodicity of the magnetic structure, it is not possible to make a direct connection between the real space configuration and spin structure factor in the reciprocal space (Fig. \ref{fig:snaps}A and Fig. \ref{fig:snaps}D represent a typical spin texture and the structure factor $S_{\bf q}^{\perp}$ respectively). To further study this phase, we introduce what we call the ``spherical snapshot'': it shows the values of the spins in the sphere where each point represents the tip of the spin centered at the origin, and the three axis correspond to the three components of the spins. 
This representation is a very useful tool in order to identify features of the spin configuration since it allows to differenciate the sublattices in a same
plot and to compare, qualitatively, the spin textures between similar or different phases. In the top panel of Fig. \ref{fig:sphesnaplowh} we show the spherical snapshots for $h/J=0.8$. Each color indicates the spins of each of the three triangular sublattices of the kagome lattice (see Fig.\ref{fig:latt}). 
Clearly, there is a symmetric distribution of the spin values in the three sublattices. This is consistent with the symmetric peak distribution in the structure factor. Interestingly, even though the inclusion of a small in-plane DM
interaction induces the emergence of non-coplanar arrangements, it is not enough to completely lift the degeneracy at low magnetic fields, leading to this 
``locally chiral'' phase.
%

\subsection{Spiral phase - $h_{c1}<h<h_{c2}$}

As the magnetic field increases, an interesting behavior is found at low temperatures.
For magnetic fields  $h_{c1} < h  < h_{c2}$, coming from the ``locally chiral'' phase described above, a spiral-like texture emerges, where there is a clear real-space  splitting in the three triangular sublattices.
This is shown in the spherical snapshot presented in the bottom panel of Fig. \ref{fig:sphesnaplowh}.
The arrangement is not symmetric: two of the sublattices are described by the same modulation with different wave vector orientation and the $S^{z}$ 
component takes all values of the unitary sphere. In the remaining sublattice (indicated by blue points), the $S^{z}$ components are restricted to positive values, with
an additional modulation. Clearly, which sublattice is arranged in which way depends on the MC realization as can be observed from the structure factor presented in Fig. \ref{fig:snaps}(E), which corresponds to a particular MC realization. However, the symmetry would be restored when averaging on several realizations, as we have checked.
This simple analysis based in the inspection of the spherical snapshot suggests that there is a sublattice symmetry-breaking induced by the magnetic field. To further explore this, and to detect the spontaneous sublattice symmetry breaking, we introduce a $\mathcal{Z}_{3}$ complex order parameter $\phi_{tot}$ defined as:

\begin{eqnarray}
\phi_{\triangle}&=&S_{1}^z + wS_{2}^z+  w^2S_{3}^z\nonumber \\
\phi_{tot}&=&\left|\frac{1}{L^2}\sum_{\triangle}\phi_{\triangle} \right|
\end{eqnarray}

\noindent where $w=\exp({i\,2\pi/3})$ and $S^z_{\alpha}$ is the $z$ component of the spins in each of the three triangular sublattices, indicated with $\alpha=1,2,3$, shown in Fig. \ref{fig:latt}.

\begin{figure}
\includegraphics[width=1\columnwidth]{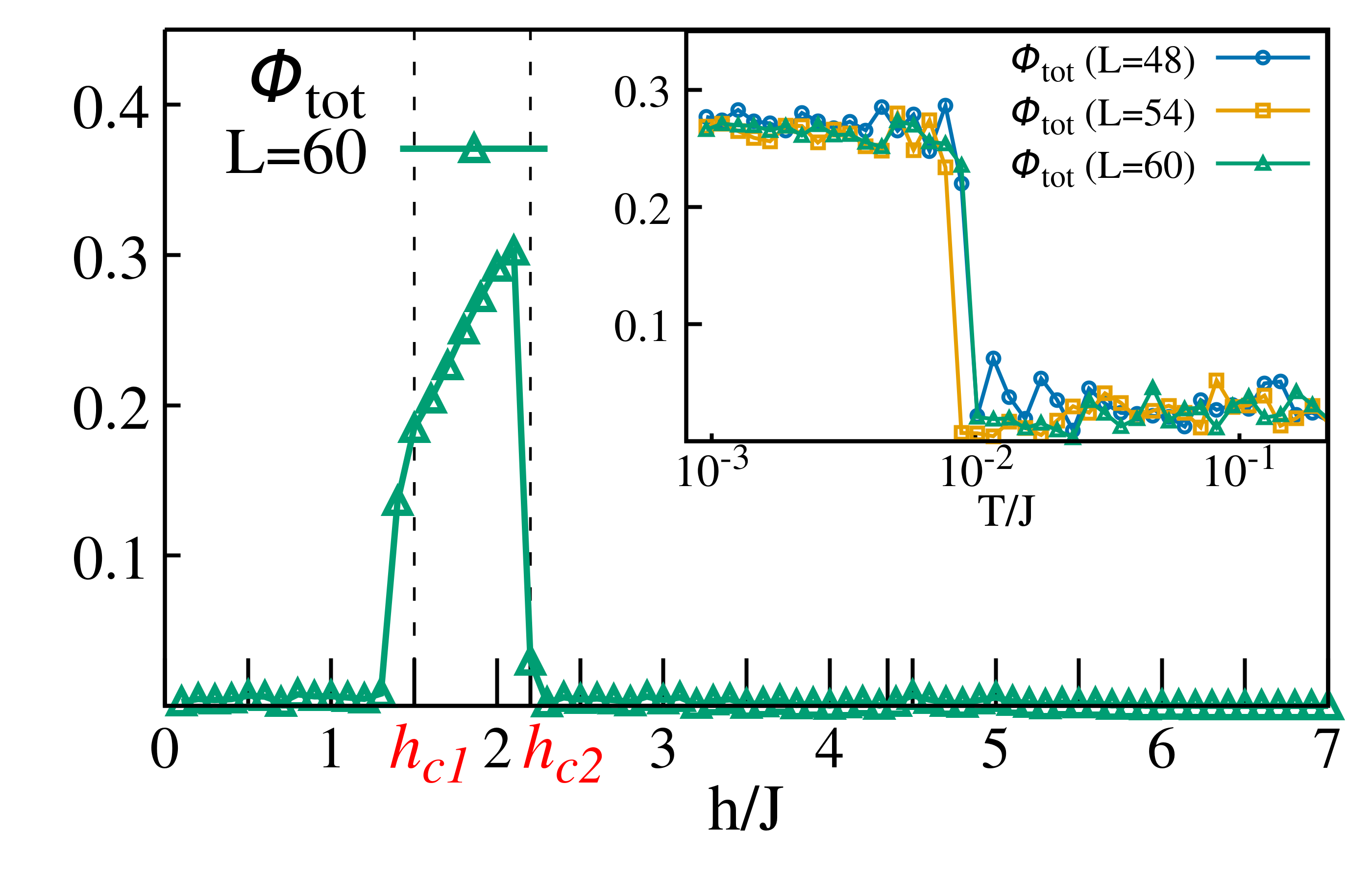}
\caption{\label{fig:mz3} (Color online) Order parameter $\phi_{tot}$ as a function of the external magnetic field at $T/J=2\times 10^{-3}$, $L=60$. 
Nonzero values of $\phi$ indicates the sublattice magnetization is not the same for each sublattice. Inset: $\phi_{tot}$ vs $T/J$ at $h/J=1.8$ for $L=48$, $L=54$ and $L=60$.}
\end{figure}

In Fig. \ref{fig:mz3} we show this parameter $\phi_{tot}$ as a function of the external magnetic field at $T/J=2\times 10^{-3}$ for lattice size $L=60$. It can be seen that this parameter is non-zero only on this spiral-like phase, where the symmetry between sublattices is broken. This feature is stable with the system size as can be seen in the inset in Fig. \ref{fig:mz3} ($\phi_{tot}$ as a function of temperature for $h/J=1.8$ and $L=48,54,60$).

%

\subsection{Antiferromagnetic pseudo-skyrmion crystal}

For $h_{c2}<h<h_{c3}$, at low temperatures a highly non trivial chiral phase emerges,
associated with a  spin texture formed by a particularly intricate three-sublattice splitting, that we call pseudo-skyrmion crystal (pSkX), very similar to the one found in the antiferromagnetic triangular lattice\cite{1}. In the triangular lattice case, the hidden skyrmion texture can be revealed by splitting the system in three interpenetrated sublattices: in each sublattice a topological skyrmion crystal SkX is stabilised. 

In the kagome lattice case, the ``pseudo-skyrmion'' name is due to the fact the these structures are similar to skyrmions, but their center is not fully polarized. As an example a typical snapshot is shown in Fig. \ref{fig:snaps} (C). 

\begin{figure}[hbt]
\centering
\includegraphics[width=0.9\columnwidth]{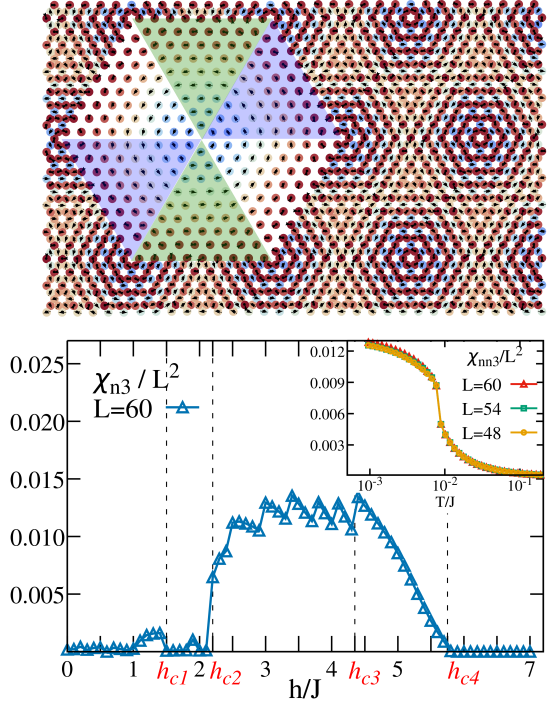}
\caption{(Color online) Top: Snapshot showing the pseudo-skyrmion structure for $D/J=0.2$, $h/J=2.6$, $T/J=2 \times 10^{-3}$. A pseudo-skyrmions is shown, erasing the neighbouring spins. Every coloured triangle indicates a diferent triangular sublattice of the kagome lattice. Bottom: Sublattice chirality density as function of $h/J$ for $L=60$ (green open triangles)
at $T/J=2\times 10^{-3}$. Inset: sublatice chirality density as function of $T/J$ 
for $L=48, 54, 60$ at $h/J=3.5$.}
\label{fig:pseudosnap}
\end{figure}

However, in the model considered here, albeit its similarities with the one proposed for the antiferromagnetic triangular lattice, the picture is not that simple. For the kagome lattice, there are also three interpenetrated sublattices of pseudo-crystals,
but these are not the three triangular sublattices constructed from the three sites in the unit cell of kagome lattice. In Fig.  \ref{fig:pseudosnap} (top) we isolate one such pseudo-skyrmion from different sublattices  at $h/J=2.6$. 
A pseudo-skyrmion is formed by an hexagon of spins joined by second nearest neighbor bonds at the center,
and it radially increases along third nearest neighbor bonds. 
The third-nearest neighbors form the three triangular sublattices of the kagome lattice. Spins belonging to each type of sublattice in the pseudo-skyrmions are   highlighted in the top panel of Fig. \ref{fig:pseudosnap}. Then, one way to extract information about the hidden structure is through the total third-nearest neighbors scalar chirality per site (i.e. the sublattice chirality) defined as:

\begin{equation}
\label{eq:chinn3}
 \chi_{n3}= \frac{1}{8\pi\,N}\sum_{\alpha=1}^3\left \langle \sum_{m=1}^{N/3} \chi_{mpq}^{(\alpha)} \right \rangle
\end{equation}
\noindent where $\chi_{mpq}^{(\alpha)}$ is the local sublattice chirality, 
 defined as Eq. (\ref{eq:local_chira}) but taking the three spins $m,p,q$ in elementary triangles in the triangular sublattices of the  kagome lattice, $\alpha=1,2,3$.
In  Fig. \ref{fig:pseudosnap} (bottom) we plot this parameter as a function of the magnetic field in for $T/J=2 \times 10^{-3}$ and $L=60$. 

\begin{figure}
\includegraphics[width=0.95\columnwidth]{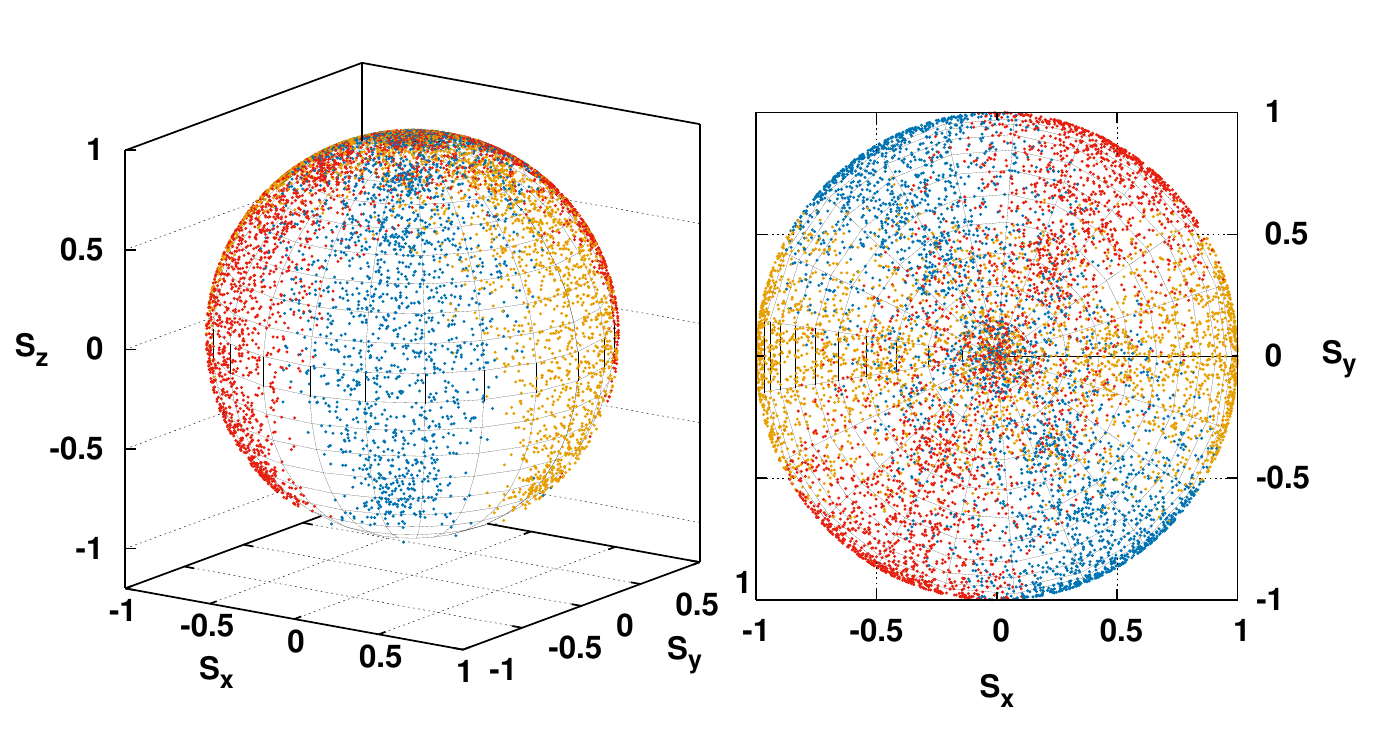}
\includegraphics[width=0.95\columnwidth]{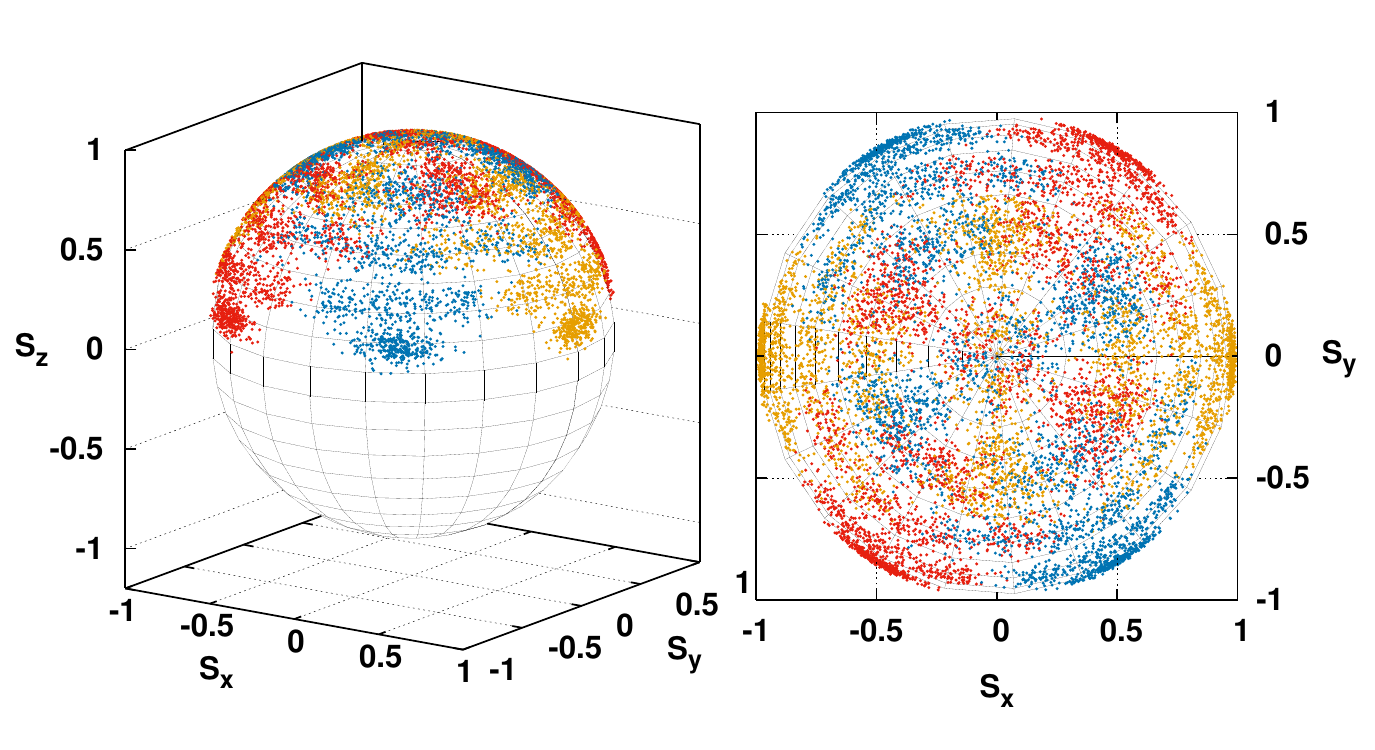}
\includegraphics[width=0.95\columnwidth]{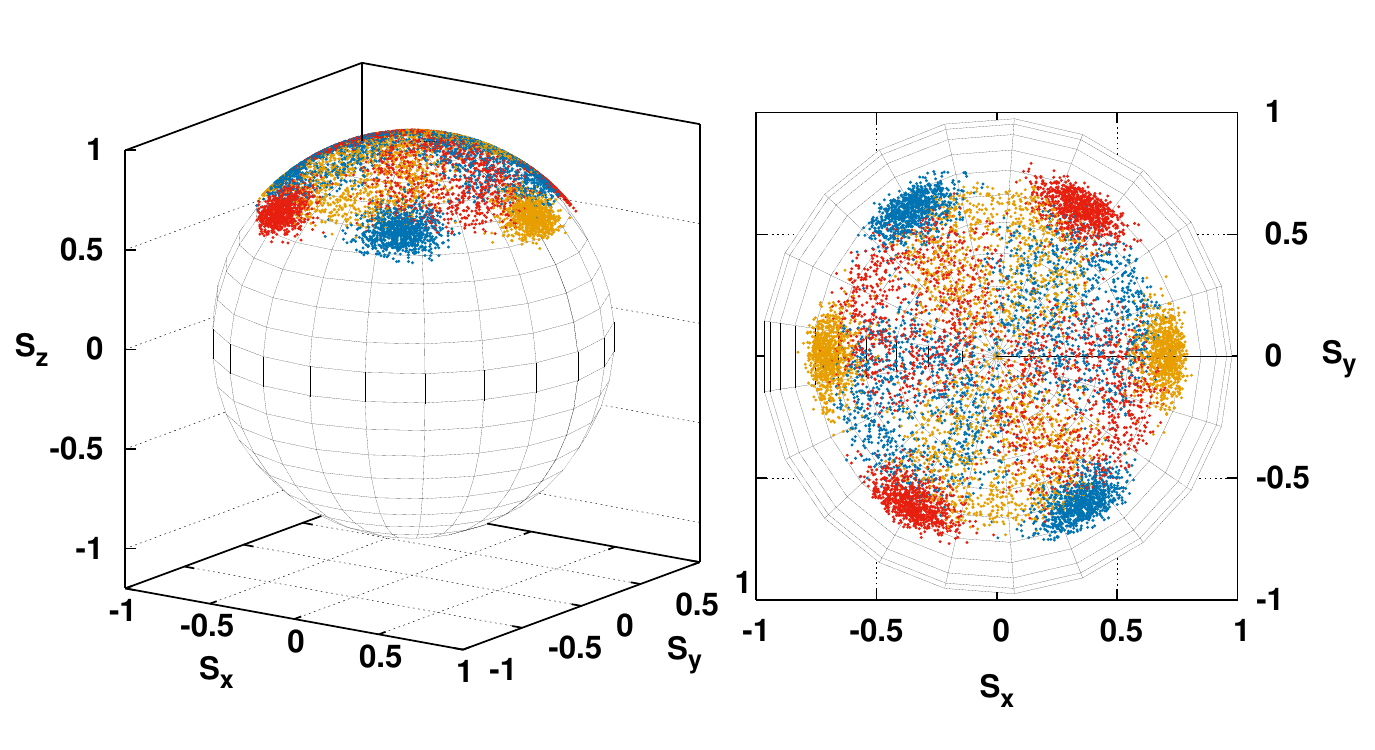}
\caption{\label{fig:sphersnappseudo} (Color online) Spherical snapshots in the pSkX phase for $L=60$ , $h/J=2.6$ (top), $h/J=3.8$ (middle) and $h/J=4.8$ (bottom) with $T/J=2\times 10^{-3}$, $D/J=0.2$. Each color indicates a different triangular sublattice.}
\end{figure}

In the low field boundary of the pSkX phase ($h/J \sim 2.5$) the local sublattice chirality $\chi_{n3}$ takes a value close to the number of pseudo-skyrmions that can be constructed.
For example, in a system with $N=8748$ sites, we found $36$ pseudo-skyrmions while the local (sublattice) chirality $\chi_{n3}\approx 31$.
This implies that the topological charge of each skyrmion is not $1$, but $Q\approx 0.86$. Hence dubbing this phase as a ``pseudo-skyrmion'' phase.
Another way to see this is that the texture associated with each ``pseudo-skyrmion'', when projected onto the sphere, does not fully wrap the sphere.
Specifically, we find that  the border of the pseudo-skyrmions is completely polarized (parallel to the field), but the core is not antipolarized, i.e.: 
the $S^z$ component never reaches the value $S_z=-1$. To illustrate this clearly, we show  a typical spherical snapshot obtained from simulations in Fig. \ref{fig:sphersnappseudo},
for three values of the magnetic field $h/J=2.6, 3.8, 4.8$, at $T/J= 2 \times 10^{-3}$ and $L=60$. As before the spins from each site of a given triangular sublattice are represented with different colors. Two significant features are 
present in this plot: the projections of the spins are divided in three ``slices'', one for each triangular sublattice, and the lowest value of the projection
along the field (found for the lowest magnetic field) is $S_z=-0.8$.

Skyrmion-like structures that do not fully cover the sphere, i.e. with $Q<1$, have already been e.g. in the anisotropic triangular lattice \cite{12} and
in \cite{25}, where there is an emergent intermediate phase between skyrmions and merons. In the model presented here, as the magnetic field increases,
the ``pseudo-skyrmion'' cores are further canted, while the edges of the magnetic structures remain parallel to the field. The evolution of the spin textures as a function of the magnetic field in 
the pSkX phase is shown in Fig.\ref{fig:sphersnappseudo}.

\begin{figure}[h!]
\includegraphics[width=0.98\columnwidth]{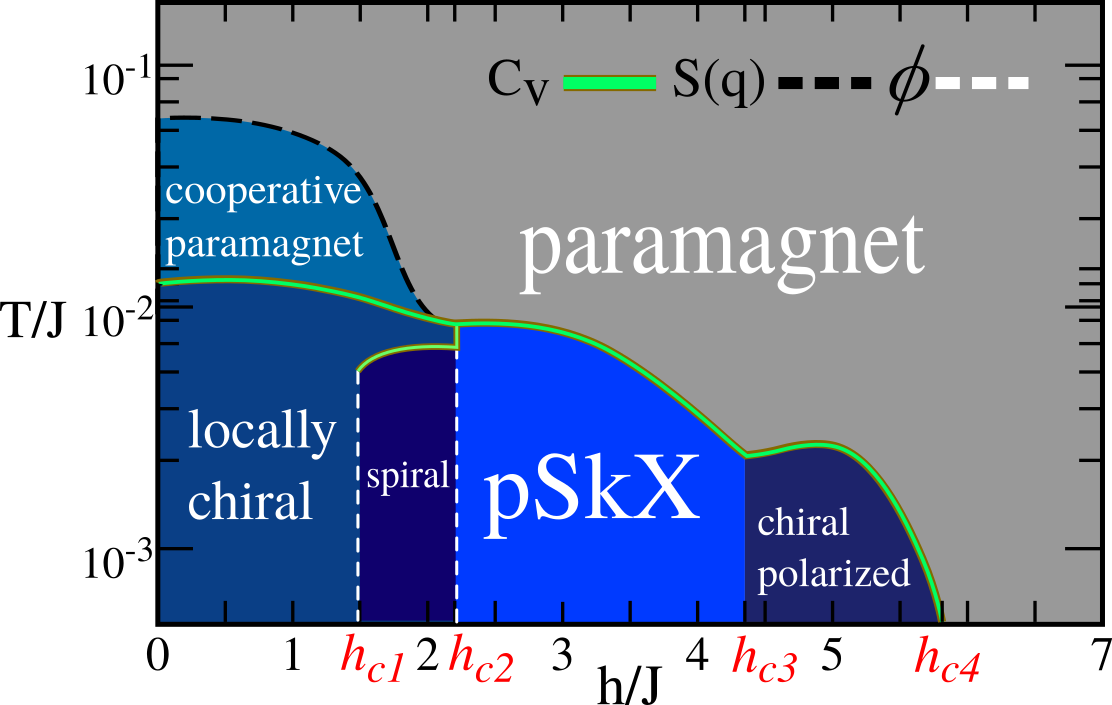} 
\caption{\label{fig:phasediag} (Color online) Complete $T/J$ vs $h/J$ magnetic phase diagram obtained from Monte Carlo simulations. The solid black line was obtained analysing  the peaks in the specific heat, the dashed white lines restrict the nonzero $\phi_{tot}$ area $h_{c1}<h<h_{c2}$, and the CP phase was obtained computing the structure factor $S_{\bf q}$.}
\end{figure}

The radius of this ``pseudo-skyrmions''
changes slightly depending on the field and system size. However, this phase is clearly present for all system sizes studied, as shown in Fig. 
\ref{fig:pseudosnap} (bottom), and it is delimited for a certain range of magnetic fields, here $ h_{c2} < h < h_{c3}$.
The sharp ``saw-tooth'' behavior of this parameter in this phase (see Fig. \ref{fig:pseudosnap} bottom) is due to the fact that the radius of the pseudo-skyrmion changes with the field.
A sharp change implies that pseudo-skyrmions with a different radii are found at that field. These magnetic structures with different radii are 
very close in energy, which explains the competition between pseudo-skyrmion crystals of different (but similar) sizes at lower temperatures.
Despite this competition, this pseudo-skyrmion crystal phase emerges and is stabilized at low temperature, 
and can be distinguished through $\chi_{n3}$, the scalar chirality calculated
in the elementary triangles of each sublattice of the kagome lattice.
No system size effects are noticed in this phase, the inset of Fig.  \ref{fig:pseudosnap} (bottom) shows $\chi_{n3}$ as a function of temperature for $h/J=3.5$ for three different system sizes ($L = 48, 54, 60$) where cleary the behavior is the same for all the cases.


In the high field region $h_{c3}<h<h_{c4}$ the spin moments are continuously further aligned with the magnetic field and the pSkX  is destroyed. Because of this, $\chi_{n3}$ decreases from its maximum value in $h_{c3}$ to zero in $h_{c4}$. We thus dub this phase the ``chiral polarized phase''.
 
With al this, combining the  $\chi_{n3}$ parameter and the information from the previous subsections, we construct the temperature vs. magnetic field phase diagram, presented in Fig.  \ref{fig:phasediag}. There are four clear low temperature phases: at low magnetic fields, there is a locally chiral phase with no clear order. In this region, at higher temperatures the system behaves like a cooperative paramagnet. As the field increases, at low temperatures, coming from the locally chiral phase, we find an intermediate spiral phase. Here, the sublattice symmetry is broken: two out of three sublattices form a complete spiral, and the third one only has positive projections along the external field. The most 
remarkable feature of the phase diagram is an extended pSkX region which is stabilized in a broad range of magnetic fields at low enough temperatures.  In this phase, pseudo-skyrmion structures are periodically arranged in three non trivial sublattices, which are in turn constructed with groups of third nearest neighbours. Therefore, we identify this phase with a topological order parameter, the third nearest neighbor chirality. As the field increases, the spins are further canted, the pseudo skyrmions are destroyed, and the chirality decreases with the field, in a chiral polarized phase.

\section{Conclusions}
\label{sec:conclusions}

The frustration in the kagome antiferromagnet is known to give rise to a plethora of exotic phenomena. The competition of different types of interactions and external fields has been shown to both relieve the frustration and induce topological phases.
In this work, we present a study  of the low temperature phases in the classical kagome antiferromagnet with competing in-plane antisymmetric Dzyaloshinskii-Moriya
interactions under a magnetic field using extensive Monte Carlo simulations. We find that, differently to what was found in previous studies in other less frustrated geometries, the particular geometry of the kagome lattice gives rise to highly non trivial magnetic orders.
Firstly, for lower fields, although the system retains some degeneracy of the pure kagome antiferromagnet, small clusters with local chirality can be identified. Interestingly, at higher temperatures, inspection of the structure factor and the specific heat shows that the system is in a cooperative paramagnet phase, as the pure kagome antiferromagnet.
As the field is increased, at lower temperatures, a three-sublattice spiral order is stabilized with broken  sublattice  symmetry: two triangular (third nearest neighbours) sublattices  form a complete spiral and in the third one the spin projection along the field only takes positive values. This allows us to construct an $\mathcal{Z}_{3}$ order-parameter, $\phi_{tot}$, to identify the extension of this phase.
Finally, we find that the external field stabilizes a pseudo-skyrmion crystal (pSkX) structure in a large portion of the magnetic phase diagram, up to the lowest simulated temperatures.
This texture is characterized by a periodical arrangement
of three interpenetrated non-trivial sublattices formed by skyrmion-like magnetic clusters. These clusters are not skyrmions, since, when projected on a sphere, the spins do not fully cover it. They have a clear polarized border and a non-fully polarized core. Moreover, due to the fact that these pseudo-skyrmions are constructed with groups of third nearest neighbors, this phase can be characterized by a topological parameter, the scalar chirality defined in each of the three 
triangular sublattices that constitutes the kagome lattice. For large enough fields, this parameter decreases rapidly to zero, as the pseudo-skyrmions are destroyed and the spins are further canted along the field.
In conclusion, we have presented and studied with extensive Monte Carlo simulations a model that combines the high geometric frustration of the pure exchange model in the kagome lattice  with antisymmetric Dzyaloshinskii-Moriya interactions, which are known to induce topologically non trivial structures when an external field is applied.  
We have found that these competing terms give rise to a rich magnetic phase diagram, where highly non trivial and topological phases are stabilized at 
low temperatures. We hope our study further contributes to the understanding of the connection between topology and frustration, where the kagome lattice 
is one of the most emblematic and relevant systems.

\section*{Acknowledgments}

This work was partially supported by CONICET (PIP 2015-813), ANPCyT (PICT 2012-1724), SECyT UNLP PI+D X792 and X788, PPID X039. H.D.R. acknowledges support from PICT 2016-4083. MV thanks Santiago Osorio for fruitful discussions.

\newpage

\end{document}